\documentclass[]{mn2e}

\usepackage{amssymb}
\usepackage{graphicx}
\usepackage{amsmath}
\usepackage{graphics}

\usepackage{color}

\title[Visible continuum and bands in the ISEC II]{On the visible continuum and bands in the interstellar extinction curve II}

\author[R. Papoular]{Renaud Papoular$^{1}$\thanks{E-mail:papoular@wanadoo.fr}\\
$^{1}$Service d'Astrophysique and Service de Chimie Moleculaire,\\
CEA Saclay, 91191 Gif-s-Yvette, France}

\begin{document}

 \maketitle
\label{firstpage}

\begin{abstract}

Any distortion of a chemical structure causes new features to appear in the absorption spectrum of the structure, especially in the visible and near UV (see Paper I). Chemical modeling, using molecular orbital theory, showed that the continuum resulting from the accumulation of weak new bands in that range, correctly mimics  the continuum measured in the laboratory on pure synthesized silicates, in the transparency spectral range, as well as in the InterStellar extinction curve in the same range. 

The present paper explores in more detail the strong discrete bands that emerge from the continuum due to distortion. It is found that different types of structure (linear or compact, carbon- or silicon-bearing) have each a limited number of strong, characteristic, bands at different wavelengths.

Distortion is only one instance of ``defects'' that enrich the vis/UV absorption spectrum; others are: vacancies and voids, substitutions, inclusions in interstices, impurities, dangling bonds. The accumulation of these result in an ``amorphous'' structural state. Several examples of known amorphous materials, both carbon- and silicon-bearing, that have been analyzed in the laboratory, and simulated theoretically, are described below, thus extending the scope of this work. A final section lists several fields of astrophysics that have used, or may use, amorphous models of dust.

\end{abstract}



In Paper I (Papoular 2019), I used molecular orbital theory to show from first principles why and how \bf a structure in its ground state, and, in particular, perfect crystal structures, in general, display no features in the visible wavelength range, and why and how imperfections (defects) in the structure give rise to bands in that range. This is because, in the ground state, all electronic orbitals below the electronic gap (separating the valence and conduction bands) are generally fully occupied, so photons can only be absorbed if they are more energetic than the gap width (barring infrared photons, which excite molecular vibrations). On the other hand, when the structure is perturbed, the orbitals are redistributed in energy and occupation so that some of the ``virtual'' orbitals above the gap become occupied, and new absorption bands appear in the spectral gap. For astronomical dust materials, they occur in the vis/near UV range. \rm

The term $defect$ designates any way the connectivity of the atoms is modified: substitution of one atom of the perfect crystal by another, inclusion of an atom in an interstice of the lattice (adatom or admolecule), vacancy or void (missing atom on a lattice node), dangling bond, deformation of the lattice cell. As the extent of the perturbation of the structure increases, features appear first in the \it transparency range \rm, between the valence and conduction bands of the perfect crystal, then increase in number and spill in the near infrared. Ultimately, the structure can be characterized as $amorphous$, and the defect  features merge into a continuum, from which only the strongest bands crop up. This process was the main subject of paper I, where it was illustrated by modeling computations on forsterite and enstatite (two varieties of silicates). That this subject is relevant to astrophysical dust is suggested by the presence of a significant $visible$ continuum in the ISEC (InterStellar Extinction Curve) and in laboratory measurements on various amorphous materials.

 In the present work, I dwell on the strength and distribution of \it discrete bands \rm of \it disordered materials \rm, and compare these properties for various carbon and silicate structures. The term disorder stresses the fact that the only type of defect considered below is \it structural deformation \rm. This is in contrast with many models of DIB carriers (see Herbig 1995), which consider heavy atomic impurities (e.g. chromium) introduced into crystals or regular molecules (e.g. C$_{60}$). It is inspired by experiments such as those of Scott and Duley \cite{sco}, who laser-synthesized forsterite and enstatite from pure Mg, Si and O, from mixtures of well defined fractional compositions. It is also reminiscent of the formation of dust by accretion in the hot winds of young stars, where crystallization of the dust structure is less likely than ``freezing'' (quenching) into an amorphous state.

This analysis of UV/vis discrete bands of amorphous structures completes the work presented in Paper I, where the continuum was studied in more detail. Together, they highlight the essential role of structural distortion in the IS carriers. However, natural and artificial amorphous materials studied in the laboratory have several other types of defects. It may therefore be helpful to give a brief account of these studies, which is done in Sec. 3

Finally, Sec. 4 explores the need and use of amorphous structures in modeling IS dust.

\section{The discrete UV/vis bands of amorphous structures}

\subsection{Computational procedure}
Here, as in Paper I, the calculations were made using the Hyperchem package, v8.0, 64 bit, released by Hypercube, Inc., and implemented on a desk-top PC equipped with a Pentium (r)II processor (450 MHz) and MMX (TM) technology. Most state-of-the-art chemical codes are available with this package, from Molecular Mechanics to various \it ab initio \rm methods. The semi-empirical PM3 code, developed by J. Stewart \cite{ste} along the same line as AM1, was used. The semi-empirical methods use a rigorous quantum-mechanical formalism occasionally combined with empirical parameters obtained from comparison with experimental results.\it PM3 incorporates a much larger number and wider variety of experimental data than AM1, and focuses on carbon- and oxygen-bearing molecules.\rm These codes compute approximate solutions of Schroedinger's equation, using some form of SCF (Self-Consistent Field) methods, such as the HF (Hartree-Fock) procedure. When the quantum mechanical calculations are too difficult or too lengthy, parameters of the method are taken from experimental data. This sometimes makes them more accurate than poor \it ab initio \rm methods, and they are always faster and can handle larger systems.

In order to allow a proper calculation of the UV/vis spectrum, one has to go beyond the SCF approximation by taking interactions between single electrons into account. The code uses the CI (Configuration Interaction) procedure for this purpose (see Foresman et al. 1992, 1995). Here the term ``configuration'' designates one distribution of available valence electrons over the orbitals. In the CI procedure, ``excited configurations'' are obtained by transfering one or more electron from occupied to unoccupied orbitals. The calculation yields a set of improved molecular states, each of which is represented by a linear combination of these configurations. This procedure is also implemented in the ``Gaussian'' chemical simulation package. For the main equations and computational details, the reader is referred to Paper I.

\bf Of course, one can also use DFT, which is now also included in these packages, to perform the same computations. 
The strength of DFT lies in its treatment of exchange interaction between electrons. In the present work, the addition of CI (configuration interaction) to PM3  also fills the need for exchange interaction (see Foresman et al. 1995, Exploring Chemistry with Electronic Structure Methods).

While DFT is probably superior for big and complicated structures, there is no indication that it beats semi-empirical codes for relatively small particles including C, H, O, N,... atoms,  precisely because the latter were empirically tailored for such common atoms. \rm

\subsection{Model structures}

\begin{figure}
\resizebox{\hsize}{!}{\includegraphics{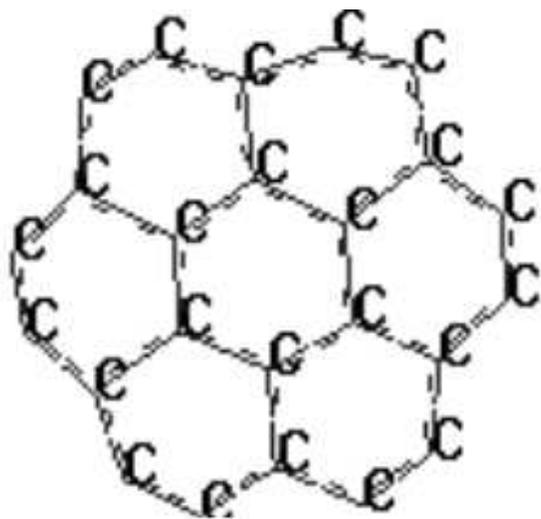}}
\caption[]{Coronene molecule C$_{24}$ in a thermal bath at 1000 K. Note the slight deformations of the hexagons; these change randomly with time; the molecule is also warped }
\label{Fig:coronene}
\end{figure}

The model structures considered here are now sketched. Figure \ref{Fig:coronene} is representative of PAHs and aromatic components of kerogene, which have been advanced as possible constituents of carbon dust. 
Figures \ref{Fig:chainC} and \ref{Fig:chainCH} are also components of kerogene (see Fig. \ref{Fig:kerog} below). Interstellar dust also contains silicates, perhaps as the dominant component. Figures \ref{Fig:forst} (Mg$_{6}$Si$_{3}$O$_{6}$H$_{4}$) and \ref{Fig:enst} (Mg$_{4}$Si$_{4}$O$_{16}$H$_{4}$) are tentative simulations of chunks of forsterite (Mg$_{2}$SiO$_{4}$) and enstatite ( (MgSiO$_{3}$)$_{2}$): the respective ratios of Si to Mg atoms are the same. 

\bf None of the structures were optimized to their ground state and all are obviously distorted and ill-terminated, as compared to their crystalline or molecular parents. This is intentional and in line with our wish to excite the samples into the amorphous state, where the virtual states above the electronic gap become partly occupied by electrons. Distortions and terminations with heteroatoms (i.e. not part of the constitutive atoms of the perfect crystals) are only instances of defects, which we are intent on creating anyway. \rm

\begin{figure}
\resizebox{\hsize}{!}{\includegraphics{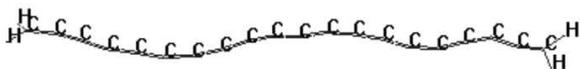}}
\caption[]{Chain of doubly conjugated (allyl) carbons, =C=C=, terminated by 2 hydrogens at each end, in a thermal bath at 1000 K. Note the undulation, by contrast with the unperturbed structure, which is linear}
\label{Fig:chainC}
\end{figure}

\begin{figure}
\resizebox{\hsize}{!}{\includegraphics{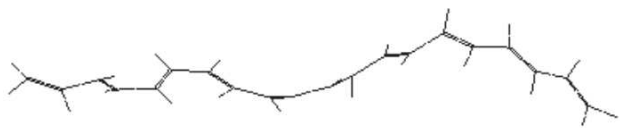}}
\caption[]{Chain of alternately singly and doubly conjugated carbons, =CH--, terminated with hydrogens at each end, in a bath at 1000 K (in line with chemistry practice, the atoms are not labeled. This is similar to $\beta$-carotene, which is known to bear a strong band about $\lambda=4 \mu$m. Here, too, the broken and undulating shape due to heating}
\label{Fig:chainCH}
\end{figure}

A disordered or amorphous structure usually comprises randomly distorted regions distributed in space within a macroscopic particle. This lack of homogeneity is simulated here by mixing the properties of 5 random realizations of each of the 5 structures, obtained by immersing the structure in a thermal bath at a high temperature and analyzing it at different successive moments. In between, the shape and orientation of the structure change. At a higher temperature, the deformation increases, as do the number of discrete UV/vis bands and their extension to longer wavelengths.

The geometrical distortions are obvious in the accompanying figures. They can be quantified by comparing the coordinates of corresponding atoms of the structure in its ground state (optimized to its minimum energy). These coordinates are computed by the code relative to the 3 principal axes of the structure. In the case of coronene, for instance, it is found the the variance of the 72 (3x24) atomic coordinates is 0.14 \AA{\ }, to be compared with a distance $\sim1.4$ \AA{\ } between adjacent atoms.

\begin{figure}
\resizebox{\hsize}{!}{\includegraphics{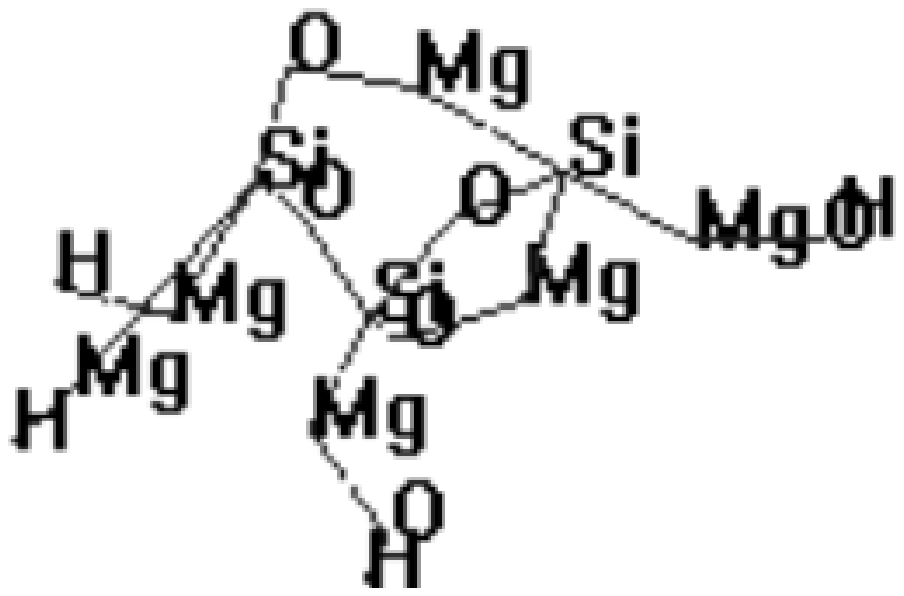}}
\caption[]{A forsterite-like structure (Mg$_{6}$Si$_{3}$O$_{6}$H$_{4}$) perturbed by heating in a thermal bath}
\label{Fig:forst}
\end{figure}

\begin{figure}
\resizebox{\hsize}{!}{\includegraphics{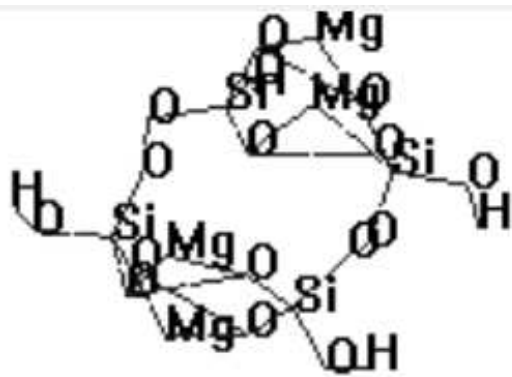}}
\caption[]{An enstatite-like structure (Mg$_{4}$Si$_{4}$O$_{16}$H$_{4}$) perturbed by heating in a thermal bath}
\label{Fig:enst}
\end{figure}

\bf The sample of structures selected in this work is only meant to visualize the notion of defects, here exhibiting geometrical distortion and chemical defects. It must be stressed, however, that the phenomena studied here are encountered, in principle, with any chemical structure, not only carbon- or silicate-rich, and with any type of defect, as made clear by the brief account at the beginning of this paper. 

Of course, a small, isolated structure distorted out of its ground state and left alone will tend by any means towards a lower, more stable state. However, if several such structures cluster together, Van der Waals forces can freeze them into a higher energy and less stable state, a precursor of an amorphous solid or a plastic polymer. \rm

\subsection{UV/vis spectra of amorphous structures}

Five oscillator strength (f)  spectra where obtained for each of the five species considered above, following the same procedure as in Paper I, and briefly recalled in Sec. 1.1. Each transition is considered to be infinitely narrow. The number of orbital configurations necessary to describe a given realization of a given species in a thermal bath varies somewhat from one to another, and so does the number of bands in the corresponding spectrum. The total number of bands (transitions) for the 5 spectra of each species is given in Tab. 1 for the 5 species: it varies from species to species, depending on the number valence electrons, and, therefore, on the chemical nature of the constitutive atoms.

\begin{table*}[ht]
\caption[]{The 5 UV/vis compiled band spectra}
\begin{flushleft}
\begin{tabular}{llllll}
\hline
Name  & Enst & Forst & coron & ChainCH & ChainC \\
\hline
Nb atoms & 28 & 19 & 24 & 42 & 24\\
\hline
Nb orbitals &100 & 64 & 96 & 102 & 84\\
\hline
Nb bands & 1194 & 1673 & 1119 & 310 & 1002 \\ 
\hline
\end{tabular}
\end{flushleft}
\end{table*}

For each species, the 5 spectra were compiled and sorted in order of wavelength. Now, it was shown in Paper I that

\begin{equation}
f=1.2\,10^{12}\frac{EW (\mathrm{cm})}{N_{a}(\mathrm{cm}^{-2})\lambda(\mathrm{cm})^{2}}=1.2\,10^{17}\frac{EW(m\AA{\ })}{N_{a}(\mathrm{cm}^{-2})\lambda(\AA{\ })^2}\,,
\end{equation}

where $EW=\tau\Delta\lambda$ is the equivalent width of the feature, $\tau$, the central optical depth, $\Delta\lambda$ the spectral width and $N_{a}$ is the column density of absorbers. By absorber is meant, here, one given computed structure.

 \begin{figure}
\resizebox{\hsize}{!}{\includegraphics{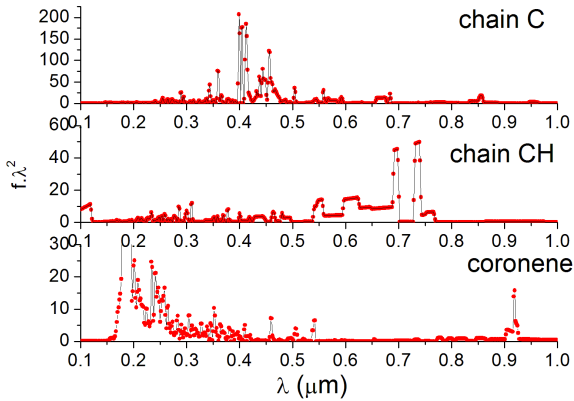}}
\caption[]{Plotted in this figure is $f.\lambda^{2}$, a quantity proportional to the band equivalent width, for the carbon model structures of Fig. 1 to 3. The interval between adjacent points is 17 \AA{\ }, so there are $\sim530$ points in all and $\sim 235$ between 0.4 and 0.8 $\mu$m. Note the sparsity of strong peaks and  the difference between peak wavelengths of different structures; also each strong peak is formed by several transitions, not only one: this shows that the oscillator strengths are not randomly distributed across the spectrum: the peaks characterize the carrier structure. Also note the large number of weak bands, constitutive of the budging underlying continuum; this is visible in the spectrum of forsterite in the next figure}
\label{Fig:Cspecs}
\end{figure}

\begin{figure}
\resizebox{\hsize}{!}{\includegraphics{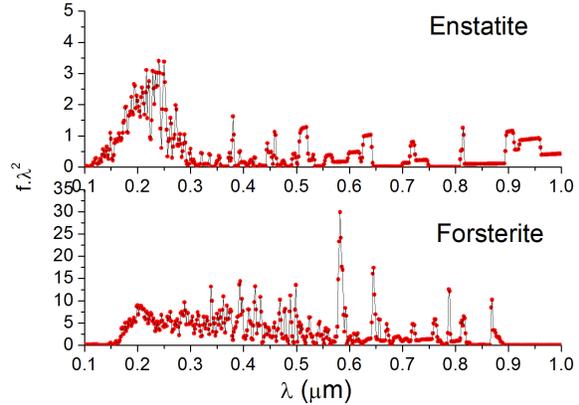}}
\caption[]{Plotted in this figure is $f.\lambda^{2}$, a quantity proportional to the band equivalent width, for the silicate-like model structures of Fig. 4 and 5. The interval between adjacent point is 17 \AA{\ }. Same remarks as in the preceding figure. The band strengths of silicate features are generally weaker than those of doubly conjugate C structures. Compact structures (coronene, silicates) exhibit UV wide bumps, hinting to conduction bands in formation}
\label{Fig:silicspecs}
\end{figure}

Figures \ref{Fig:Cspecs} and \ref{Fig:silicspecs} plot $f (\lambda)\lambda^{2}$, a quantity proportional to the equivalent width, for the 5 species. Each spectrum is drawn as a black line joining adjacent dots. The interval between adjacent points is $\sim17$ \AA{\ }. Several remarks can be made about these spectra:

- they differ from species to species;

- for a given species, only a few strong peaks emerge beyond 0.3 $\mu$m, from a sea of much weaker ones;

- these peaks differ in wavelength, from one species to another;

- the $f.\lambda^{2}$-level of the underlying quasi-continuum is of order 1;  

- each of the strongest peaks covers a few dots, indicating that  the strong transitions are not distributed randomly;

- compact structures (coronene, silicates) show increased absorbance over the UV range (below 0.3 $\mu$m),
a region corresponding to the conduction band of solids, where absorbance is known to be very strong.

- the amorphous structures which contribute to the strongest bands are the doubly  conjugate chains, forsterite and the singly conjugate chains.

\it Thus it seems that, while the number of transitions increases indefinitely with the size of the carrier, the strong ones flock around a finite number of wavelengths, characteristic of the given structure. Transitions rarely appear beyond the near-IR. \rm 

\begin{figure}
\resizebox{\hsize}{!}{\includegraphics{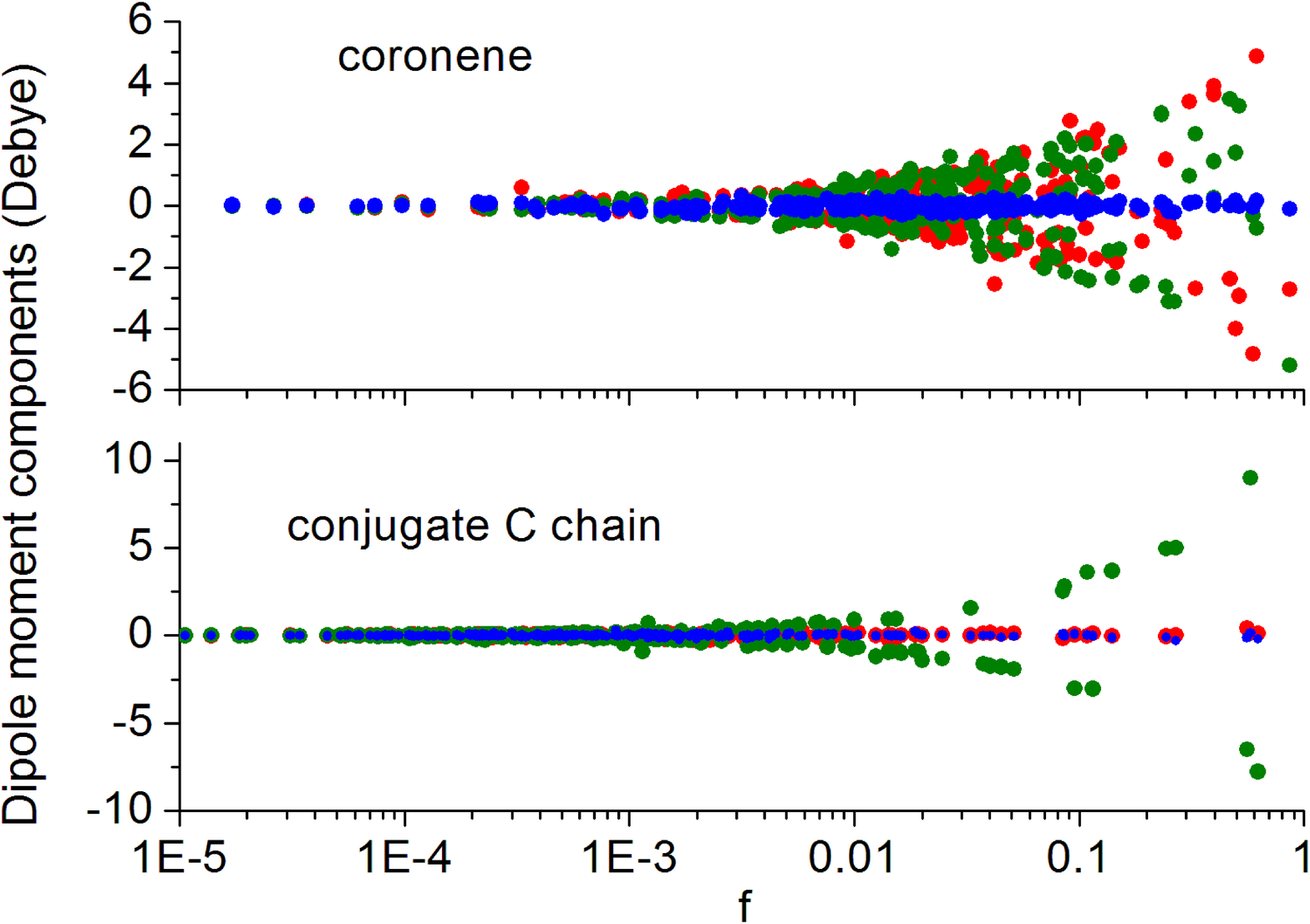}}
\caption[]{The components of the transition electric dipole moments of a linear (C chain) and a compact (coronene) planar structures, against the corresponding oscillator strengths. They lie along the respective principal axes of the structures. In the compact structure, the dipoles are randomly oriented in the coronene plane; in the linear structure, they are all along the backbone of the chain}
\label{Fig:dipmomts}
\end{figure}

Still another remarkable feature of these figures is the particularly strong, but few, bands of the doubly and singly conjugate chains. This becomes understandable if it is remembered that the oscillator strength scales like the square of the electric dipole moment of the corresponding transition: obviously, a nearly linear structure favors a much higher moment than a compact one, like coronene or silicates. This is illustrated in Fig. \ref{Fig:dipmomts}.

Comparison of wavelengths and strengths with those in Paper I suggest that the wavelengths are less sensitive to distortion than are the strengths.

It is remarkable that nearly all the spectral characters mentioned above are logical consequences of the bands originating in the random distribution of the electronic orbitals of disordered structures rich in electrons, as developed in Paper I.
 
\section{DIBs and the UV/vis discrete bands}
Since, on the one hand, DIBs (Diffuse Interstellar Bands) have been detected in large, and still growing, number ($>500$; see Hobbs et al. 2008, 2009), and, on the other hand, assuming dust is in amorphous form, discrete bands necessarily appear in the transparency range, it is tempting to compare both groups in wavelength and intensity. However, the accuracy of the modeling calculations reported here has not been ascertained, so individual identifications are excluded; even so, the fact that these calculations use an \it ab initio \rm code suggests that the comparison still makes sense, at least in a qualitative sense.  

\begin{figure}
\resizebox{\hsize}{!}{\includegraphics{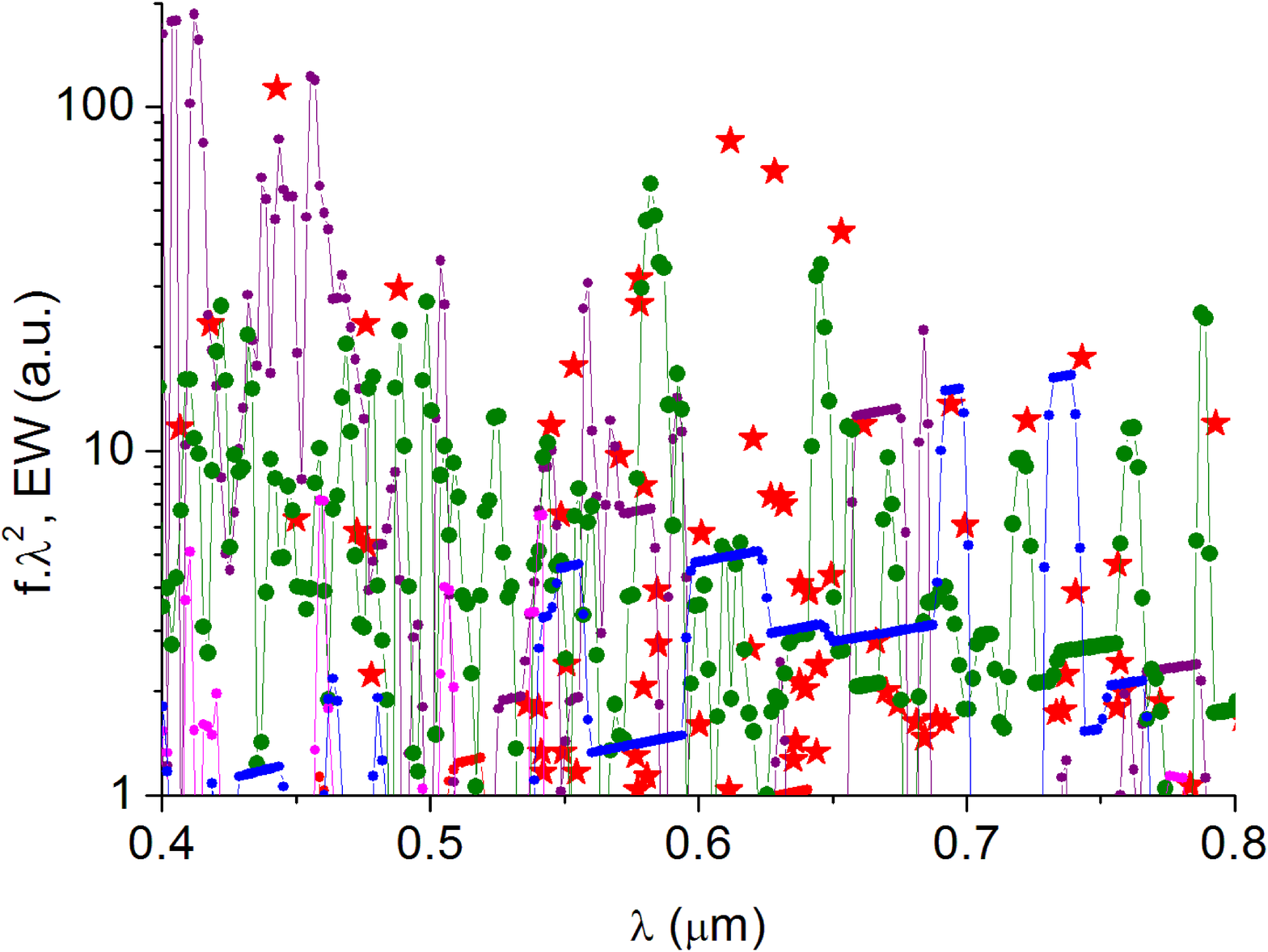}}
\caption[]{\it black stars: \rm equivalent width, $EW\,$(m\AA{\ })/30, of the 127 DIBs as tabulated by Herbig \cite{her}.
Also plotted ($dots$) are each of the 5 spectra (1200 points) in Fig. \ref{Fig:Cspecs} and \ref{Fig:silicspecs}, multiplied in intensity by a  roughly tailored constant, C. For enstatite ($red$), forsterite ($olive$), coronene ($magenta$), chainCH ($blue$) and chainC ($purple$), the corresponding constants are: C= 1, 2, 1, 1/3, 1 (see comments in text); the interval between adjacent model bands is 17 \AA{ }}
\label{Fig:comparison}
\end{figure}

Figure \ref{Fig:comparison} plots first the  scaled equivalent width, $EW\,$(m\AA{\ })/30, of the 127 DIBs as tabulated by Herbig \cite{her} (black stars). Each of the 5 spectra (1200 points) in Fig. \ref{Fig:Cspecs} and \ref{Fig:silicspecs}, multiplied in intensity by a  roughly tailored constant, C, is also plotted. For enstatite (red), forsterite (olive), coronene (magenta), chainCH (blue) and chainC (purple), the corresponding constants are: C= 1, 2, 1, 1/3, 1.  

The most striking characters common to both families of bands are 

- their flocking in large numbers in the visible and near-UV ;

- the wide ranges of their equivalent widths, most of them being weak and narrow;

- the weakest bands being the more densely packed;

- the few strongest bands near 4400 and 6000 \AA{\ };

- their absence beyond the near-IR;

- the DIBs are known to be only weakly polarized, as are the compact distorted structures studied above (see Fig. \ref{Fig:dipmomts}); the carbon chains are weakly polarized, too, unless they are strictly linear; polarization is further reduced for grains formed by random coagulation of several such structures:

- the model bands that emerge from the underlying continuum are found to be a compendium of several discrete transitions. These transitions are legion and seem to be randomly distributed in wavelength; consequently, the width of the compounded bands varies widely. Similarly, DIB widths vary from less than 1 up to tens of Angstroms ; many wide DIBs, each initially considered as a single transition, have later been spectrally analyzed into several close transitions. 

- obviously, under these circumstances, the model bands can rarely be symmetric in profile; this is also the case for the DIBs (see Krelowski 2008).

- since each species/structure carries a number of characteristic bands, there is a correlation between the intensities of these bands; if absorbing clouds harbor grains of different species/structures with the same fractional composition, then all band intensities will be correlated. To some extent, this is also the case of DIBs. In the present band model, these are carried by the same grains that carry the IS absorption continuum (see Paper I), so the band and continuum intensities must be correlated, as is observed, to some extent, between DIBs and E$_{B-V}$.

The column densities of each type of absorbers can be deduced from the spectra, using Eq. 1: $N_{a}=$ 4, 2, 4, 12 and 4 $10^{15}$ cm$^{-2}$. These numbers are an order of magnitude higher than the estimate of Paper I, mainly because the structures are not the same and are less defective than those of Paper I.
 
\it It is apparent, therefore, that, if indeed DIBs are carried by distorted chemical structures, then DIB strengths will depend on the degree of perturbation of the carrier structures away from their optimum configurations  (strongest binding energy), and so may vary from parent cloud to parent cloud, in agreement with Krelowski's statement that all DIBS are variable \cite{kre}. Comparison with the results of Paper I also indicates that the penetration of the bands into the IR is limited by the extent of the distortion \rm.

This study of the discrete features in the UV/vis spectrum of distorted structures completes the work begun with the study of their underlying continuum in Paper I. However,  structural disorder is not the only way to create such features. The next section shows that natural and artificial amorphous materials studied in the laboratory carry several other types of defects, and produce similar spectra.

\section{Characterization of the amorphous state}
The amorphous state is best defined by contrast with the crystalline state. In the latter, atoms of a small number of types are regularly positioned at the nodes of a highly symmetric lattice. In particular, it has translational symmetry: a translation of an integral number of lattice periods in any direction superposes the lattice on itself. In this state, the energy binding the atoms is maximum and the potential energy, minimum. This can be realized only under  very special circumstances, which are not common: geometric regularity is easily impaired by the presence of foreign (hetero) atoms, or any other $defect$. Defects come in various categories defined by the way the connectivity of the atoms is modified: substitution of one atom of the perfect crystal by another, inclusion of an atom or molecule in an interstice of the lattice (adatom or admolecule), vacancy or void (missing atom on a lattice node), dangling bond, deformation of the lattice cell.

\begin{figure}
\resizebox{\hsize}{!}{\includegraphics{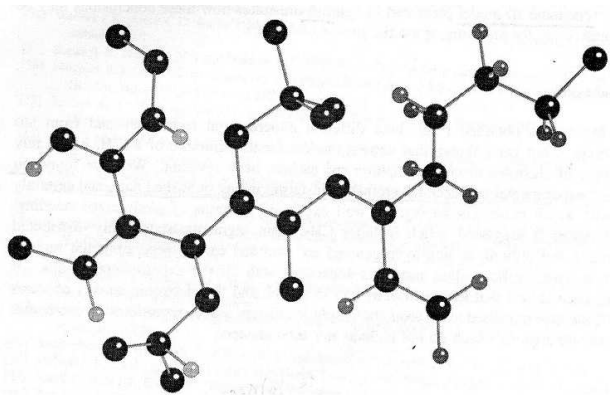}}
\caption[]{Schematic model of the microstructure of a-C:H (amorphous hydrogenated carbon, also designated by HAC), adapted from Walters and Newport \cite{wn}. The larger, black balls are C atoms; the smaller, gray ones are H atoms. There are CH, CH$_{2}$ and CH$_{3}$ groups, and sp$^{2}$ and sp$^{3}$ bonds.}
\label{Fig:hacwn}
\end{figure}

\begin{figure}
\resizebox{\hsize}{!}{\includegraphics{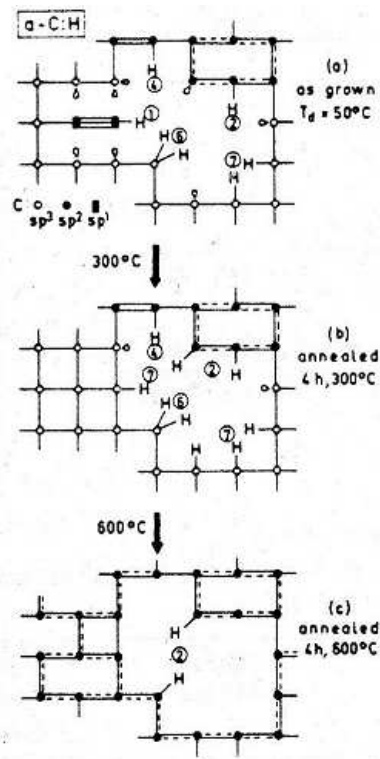}}
\caption[]{Proposed model for the thermally induced changes in a carbon-rich microstructure, adapted from Fig. 3 of Dischler et al. \cite{dis83}. Note the 4- and 6-membered rings as well as microvoids}
\label{Fig:hacdischl}
\end{figure}

Figures \ref{Fig:hacwn} and \ref{Fig:hacdischl} are examples, taken from the literature, of representation of an amorphous material obtained by perturbing a graphite structure in several different ways. 

\begin{figure}
\resizebox{\hsize}{!}{\includegraphics{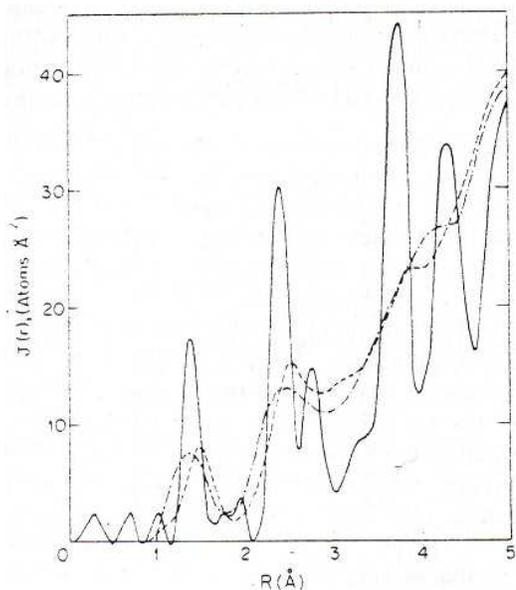}}
\caption[]{Experimental radial distribution functions of glassy carbon (solid curve) and two different evaporated carbons (dashed and dash-dotted curves), adapted from Fig. 4 of Robertson \cite{rob}. The radial distance, R, is counted from an arbitrarily chosen C atom. Note the large, distinct peaks in the former case (regular lattice) and, in the latter (amorphous samples), the weaker peaks of decreasing amplitude as the radial distance increases. }
\label{Fig:raddistr}
\end{figure}

The degree of amorphization depends on chemical composition, preparation, heat and pressure applied, and age. As the density of defects increases the system looses its regularity and, in the limit, the lattice becomes a $random$ network and the body becomes completely $amorphous$. One way to illustrate the difference between the two limiting states of solid matter is to plot the number of atoms found at a given distance from a reference node, as a function of that distance (the \it radial distribution function\rm). An example of such a plot is reproduced in Fig. \ref{Fig:raddistr}, from Robertson (Fig. 4, 1986): distinct, strong peaks are apparent in the case of crystalline materials; for amorphous materials, they are reduced to small ripples.

\begin{figure}
\resizebox{\hsize}{!}{\includegraphics{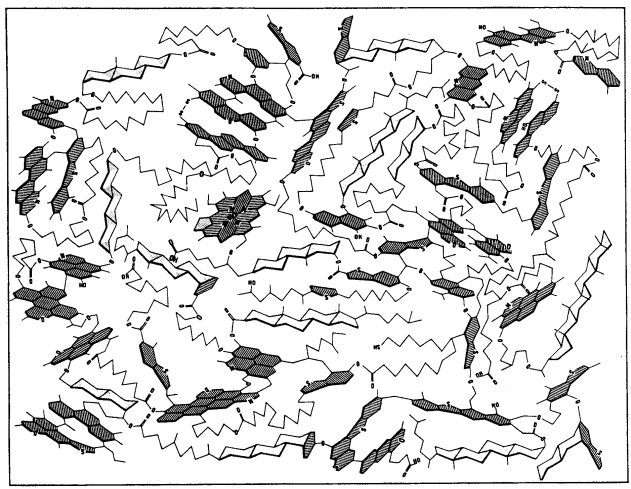}}
\caption[]{Chemical representation of a type II kerogen, adapted from Fig. 3 of Behar and Vandenbroucke (1986). Following common practice, C-H bonds are not represented. Carbon chain skeletons are shown as broken, undulating lines. The aromatic clusters of benzenic rings are shaded. Various functional groups and oxygen bridges are labeled.}
\label{Fig:kerog}
 \end{figure}

Examples of naturally amorphous material found in earth are kerogens and coals. Apart from C and H, these contain a large fraction of oxygen and small fractions of N, S, etc., which relates them to the so-called CHONS  and MAONs families introduced in the study of IS dust (see Kwok and Zhang 2011). They evolve slowly in time and with depth in earth, from less to more crystalline forms.
A typical representation of these materials is reproduced in Fig. \ref{Fig:kerog} . Note the lack of homogeneity, the aliphatic and aromatic clusters of benzenic  rings, the undulating carbon chains and the oxygen bridges in between.

\begin{figure}
\resizebox{\hsize}{!}{\includegraphics{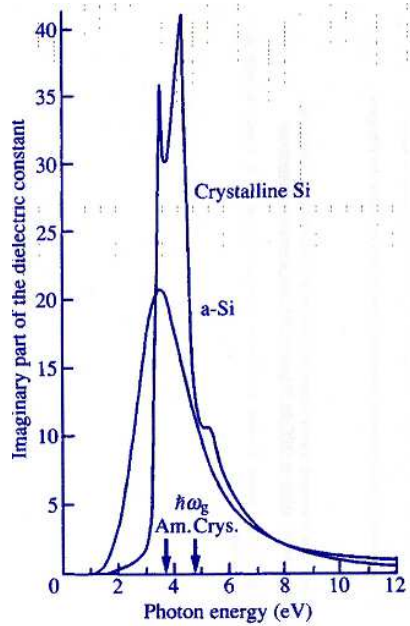}}
\caption[]{Imaginary part, $\epsilon_{2}$, of the dielectric function fpr a-Si and crystalline silicon (from Pierce and Spicer  1972, reproduced in Street 1991). The photon energies at the peaks are marked with vertical arrows.}
\label{Fig:a_Si}
\end{figure}

\begin{figure}
\resizebox{\hsize}{!}{\includegraphics{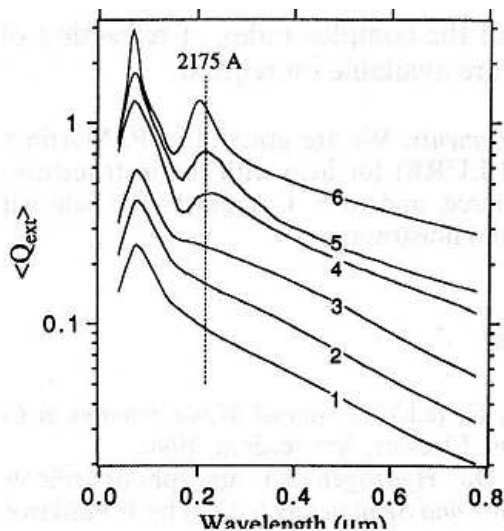}}
\caption[]{Optical extinction efficiencies for small grains of coal of increasing degree of coalification (collected from increasing depth in Earth, and hence increasing degree of graphitization), curves 1 to 4. For comparison, the same for Polycrystalline Graphite (5) and Highly Oriented Pyrolitic Graphite (6). Adapted from Fig. 2 of Papoular et al. \cite{pap95}. In all case the grain size distribution is in $a^{-3.5}$, with $a_{min}=20$ \AA{\ } and $a_{max}=2500$ \AA{ }. The peak at 2175 \AA{\ } is due to $\pi-\pi^{*}$ electronic transitions and the peak farther into the UV, to $\sigma-\sigma^{*}$ transitions. Note their progressive waning from top to bottom as the material becomes more and more amorphous.}
\label{Fig:coalevoln}
\end{figure}

Spectroscopically, macroscopic amorphous bodies differ from crystals in that their characteristic bands tend to be wider and, generally, weaker as illustrated in Fig. \ref{Fig:a_Si}  for crystalline and amorphous Si (Street p. 86, Fig. 3.17).  Fig. \ref{Fig:coalevoln} shows the corresponding spectral evolution of aging coal in the near-UV/visible range: the characteristic graphite 2175 \AA{\ } feature is seen to increase in strength and decrease in width from bottom up, as the material is buried deeper in earth and grows older. This annealing process is called coalification.

\begin{figure}
\resizebox{\hsize}{!}{\includegraphics{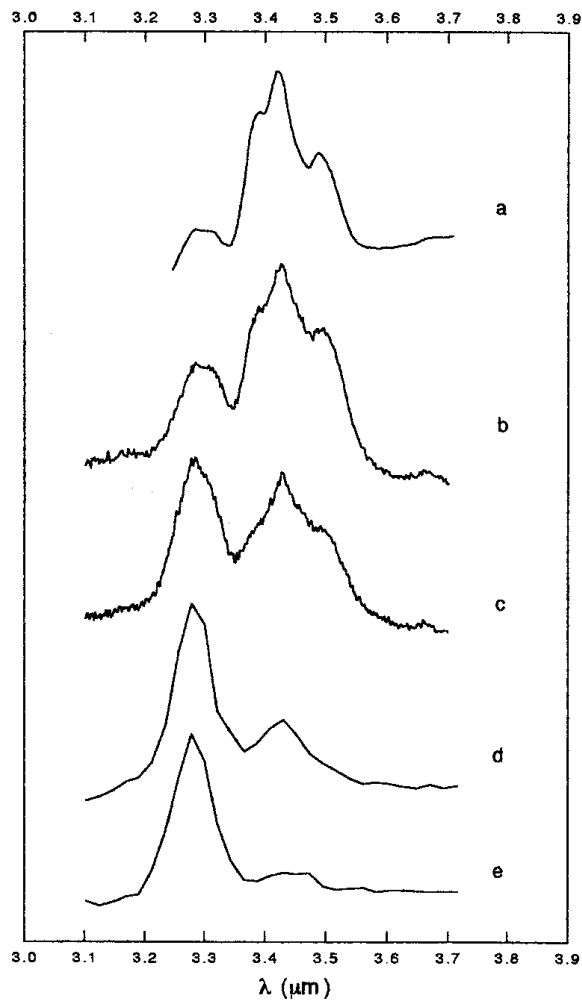}}
\caption[]{Laboratory C-H stretch band of coals for increasingly higher evolutionary stage, from a to e: decreasing ratios H/C and O/C, increasing graphitization. Adapted from Guillois \cite{gui}}
\label{Fig:CHstrcoal}
\end{figure}

\begin{figure}
\resizebox{\hsize}{!}{\includegraphics{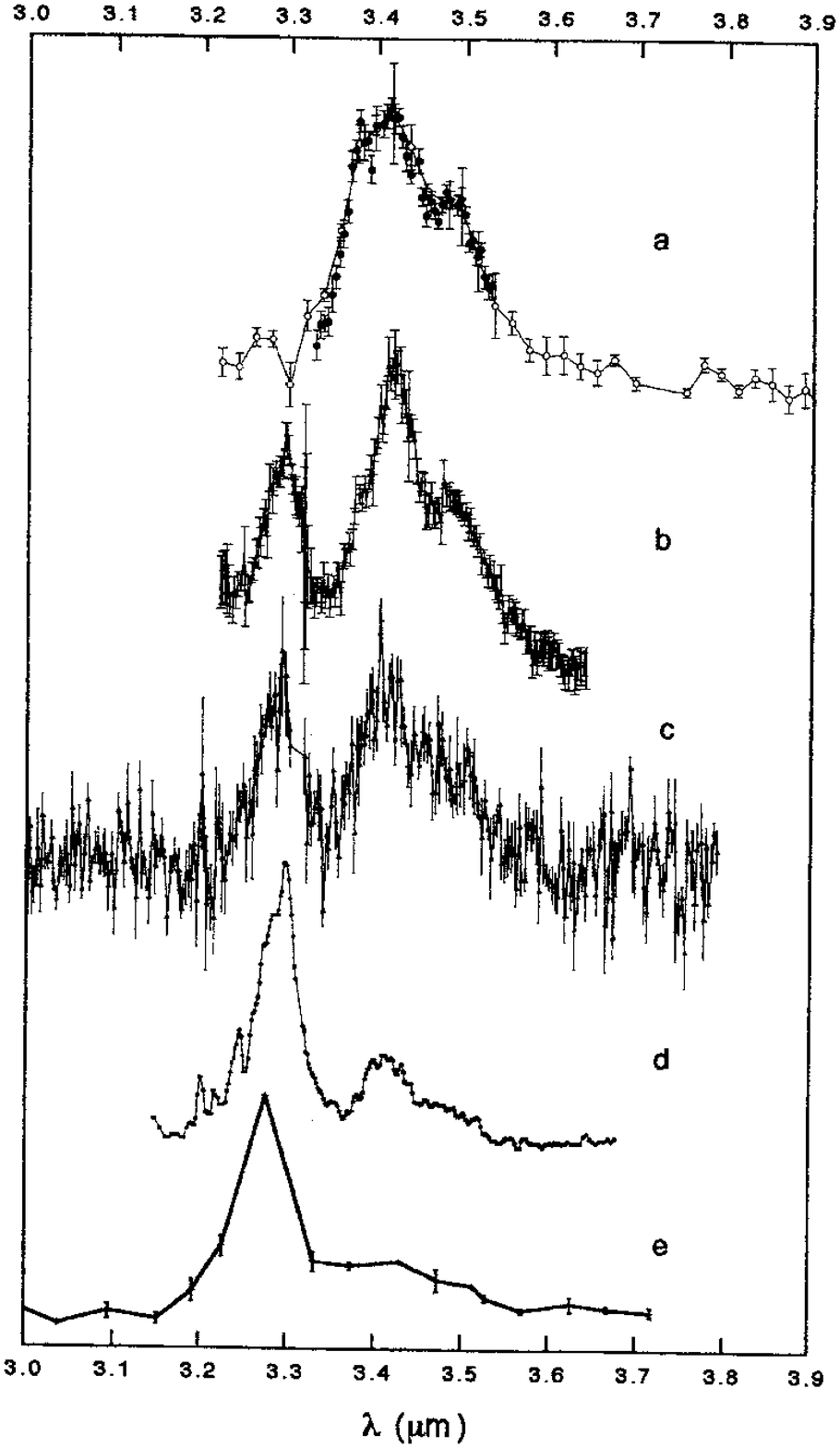}}
\caption[]{The astronomical C-H stretch band of a) GC/IRS6E in absorption; b-d) 3 post-AGB stars, in emission; e) reflection nebula NGC 2023. Adapted from Fig. 5 of Papoular \cite{pap01}, where the journal sources are available. Note the similarity of these plots with the corresponding ones of Fig. \ref{Fig:CHstrcoal}}
\label{Fig:CHstrIS}
\end{figure}

In view of the complex structure and lack of homogeneity of coal and kerogen, it comes as no surprise that their spectral  features are often composite. The C-H stretch band, for instance, is a variable superposition (multiplet) of overlapping features traceable to various aromatic (sp$^{2}$) and aliphatic (sp$^{3}$) functional groups. The relative strengths of the corresponding bands varies along the annealing process, as seen in Fig. \ref{Fig:CHstrcoal}. In the upper spectrum, the aliphatic peaks dominate near 3.4 $\mu$, in the lowest the aromatic peak near 3.3 $\mu$ is dominant. Anticipating on the next Section, Fig. \ref{Fig:CHstrIS} shows a parallel evolution observed in the IS medium.

The causes of band broadening are many, even barring superposition of overlapping bands (see Papoular 1999). Precisely because of the disorder in their structure, amorphous solids are particularly subject to \it inhomogeneous broadening\rm, due to the differences of environments surrounding a given group of atoms carrying the band considered. For small grains with a given amorphous structure, this broadening is bound to decrease with the grain size, as does the relative number of bulk atom groups compared with surface groups. It also depends on whether the grain is excited  in the bulk or only at the surface (for example, by atomic impact).

These effects are compounded with the positive correlation of width with size according to Mie's theory (see Draine and Lee 1984).

\begin{figure}
\resizebox{\hsize}{!}{\includegraphics{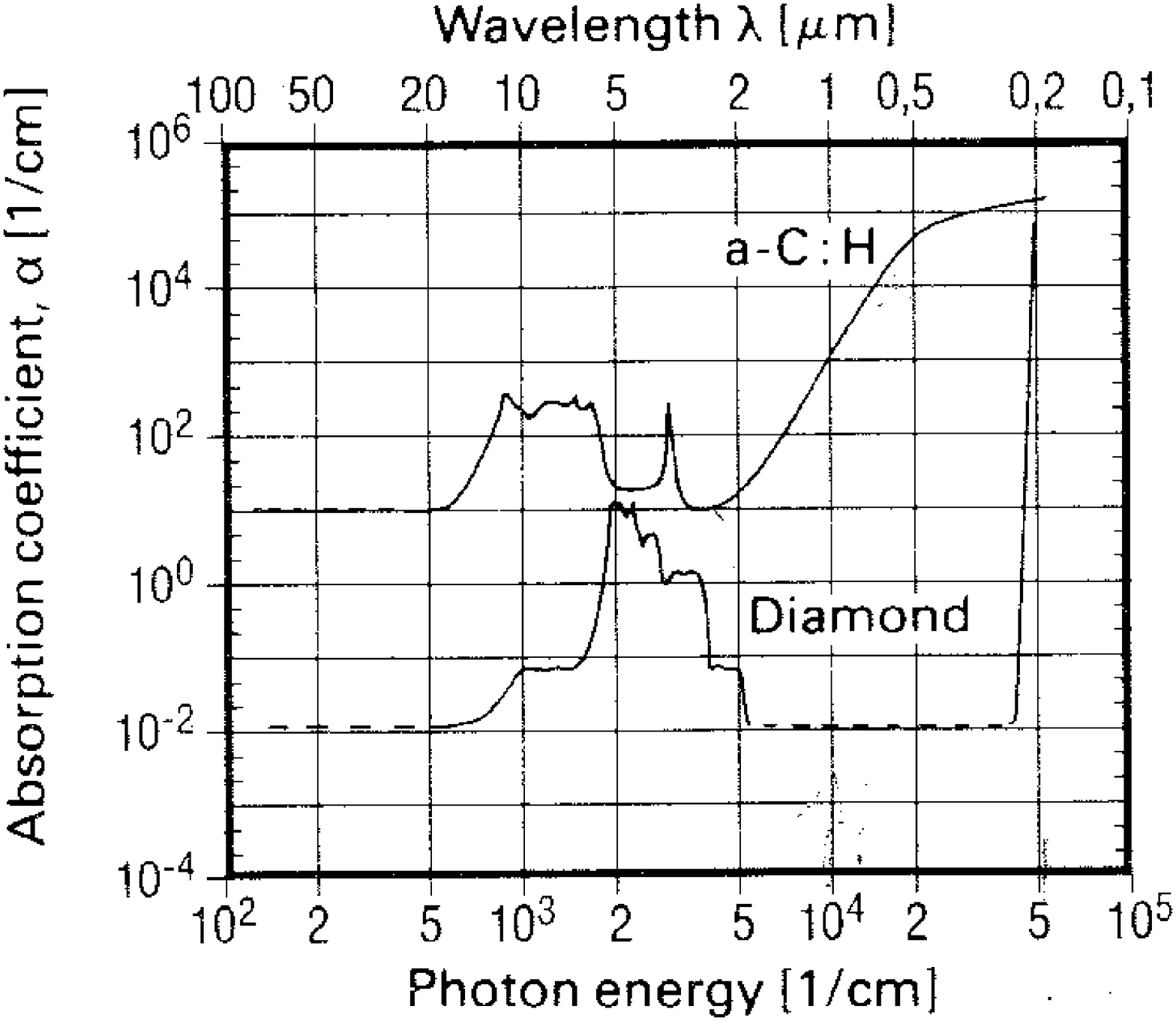}}
\caption[]{Spectra of absorption coefficients of a-C:H and diamond, adapted from Fig. 2 in Dischler and Brandt \cite{dis85}. Note the a-C:H continuum all over the spectrum, but increasing steeply below 2 $\mu$m, by contrast with the very low continuum of diamond down to the UV.}
\label{Fig:haccontinuum}
\end{figure}

Another remarkable spectral characteristic of amorphous materials is their bearing an extensive continuum underlying the bands. Figure \ref{Fig:haccontinuum} taken from Dischler and Brandt \cite{dis85} displays, for a particular sample of a-C:H (upper plot), a notable continuum increasing steadily from the far IR to the near UV. This is not the case of diamond (lower plot). In the infrared, such continuum  is likely due to excitation of lattice phonons by the vibrating functional groups and by incident light. The maximum phonon frequency scales like the total number of atoms in the sample and extends into the near IR (see Kittel 2004). The continuum at higher wave numbers is much stronger and mainly due to structural defects and impurities, as detailed in Sec. 2 and Paper I. By contrast, in the visible and near UV, many crystals display a \it transparency range.\rm 

While crystal structures are naturally determined in finite numbers by their composition, amorphous structures also depend on the way they are produced. This seriously complicates the choice of amorphous models for IS dust. It makes it necessary to compare various types of amorphousness, both experimentally and theoretically.

The characters of amorphous structures also partly depend on their chemical composition. Here, we are interested in the main components of IS dust: carbon-rich and silicate materials. Typically, laboratory models are obtained in the amorphous state by mixing the component elements in the right proportions into a solid pellet. Upon irradiation of the latter with a powerful laser in an argon atmosphere, the solid is vaporized and its ejecta condense on a slide in the amorphous state. The structural complexity of the naturally amorphous kerogens and coals is an example of the combined effects of composition and synthesis processes.

The potential energy of an amorphous system is higher than that of the crystal of the same chemical composition. This is reminiscent of Walsh's rules for molecules: in general, in the ground state of the molecule
(electric neutrality, geometric symmetry  and maximum binding energy) , its potential energy is minimum; Walsh's rules compile the increase of the molecular potential energy above the ground state energy, for a number of typical deformations of the structure.

The modeling results presented in this and Paper I, together with previous experimental results described in this section,  contribute to foster amorphous structures as important constituents of IS dust, a subject taken up next.

\section{The need for amorphous dust models}
The notion of dust in astrophysics was introduced by Trumpler \cite{tru} nearly a century ago in order to explain observed absorption and scattering of star light. At that time, the theoretical understanding of the solid state was still in its infancy, following the ongoing developments of Quantum Mechanics. It was more straightforward to apply its tools and methods to crystals than to disordered material. Understandably, then, the notion of amorphous solids is hardly mentioned in the literature of the epoch, although most common materials, of course, are not crystalline. For simplicity, however, the small IS dust grains were mostly treated as if they were made of a perfectly random atomic lattice, with no explicit mention of it: thus, the essential optical quantity, the index of refraction was, simply prescribed in the form $m=n+ik$. With the Mie theory at hand, the grain size was the main tailoring parameter (e.g. see van de Hulst 1957); others were the chemical composition (ice, graphite, silicates, silicon carbide, etc.) and the thickness of ice coatings over cores of other materials. The anisotropy was explicitly treated mainly in the case of graphite which was deemed necessary to explain the 2175-\AA{\ } feature(e.g. see Draine and Lee 1984). Great strides in understanding the ISEC were made under these circumstances, leading to a number of dust models being proposed. One of the most popular was developed by Mathis and collaborators (see Mathis and Wallenhurst and literature cited therein, 1981).

By mid-past century, electronic technology became interested in amorphous materials, especially derivatives of silicon and carbon. The essentials of the properties of such materials can be found in Mott and Davis \cite{MD}, Robertson \cite{rob} for amorphous carbon, Street \cite{str} for silicon, for instance. At the same time, the development of infrared technology allowed astronomers to explore the IR sky and, in particular, detect the silicate band at 9.7 $\mu$m. Huffman \cite{huf} and Day and Donn \cite{DD} initiated the study of amorphous dust candidates by condensing amorphous  grains of different compositions, such as magnesium silicate (Mg$_{2}$SiO$_{4}$) smoke particles from hydrogen and argon atmospheres containing Mg and SiO. Heating to 1000$\char'27$C converted them into the crystalline form.

About the same time, Sagan and Khare \cite{sag} and Gradie and Veverka \cite{GV}, studying the composition of asteroids, introduced the concept of $tholin$, a complex organic solid matter similar to aged kerogen (see previous Section).

Following these developments, Mathis and Whiffen \cite{MW} argued that their models were improved by using 
optical properties of amorphous materials rather than the crystalline allotropes. In the case of carbon dust, the discovery of the Unidentified Infrared Bands (UIB), from 3 to 13 $\mu$m, led Duley and Williams \cite{DW} to draw attention to the chemical reactivity of carbon and build a model in which the grain surface carries hydrogen functional groups (CH, CH$_{2}$, CH$_{3}$), each of them providing a band in the near to mid IR; they assumed the bulk grain material to be amorphous. 

Yet, later models favored crystalline structures, presumably due to the strength, anteriority  and ubiquity of the 2175 \AA{\ } feature and to the availability of measured and computed band spectra of polycyclic aromatic hydrocarbons (PAH). These are \it planar, aromatic, fused rings molecules (as opposed to bulk material) \rm, including only C and H, the latter in the form of functional groups at the periphery of the molecules (as in Duley and Williams, 1981). Fitting the observed spectra required adding the spectra of a large number of molecules of this family, adding atoms other than C and H, such as O and N, and increasing the particle sizes (Chiar and Pendleton 2008).

Even so, the observed bands remain unusually wide for molecules like PAHs, even if rotation broadening is taken into account. More importantly, the inception of dust in space, by random accretion of gaseous atoms from a very rich and diversified  bath,  make it quite unlikely that the outcome could be a highly symmetric molecule, let alone a crystalline grain. In the laboratory, such a result is only obtained by high-temperature processing. In star ejecta, on the other hand, dust grains appear to condense below 1000 K.

These experimental and observational results led to the development of several amorphous dust models, most of them based on Hydrogenated Amorphous Carbon (HAC or a-C:H) (see  Williams and Arakawa 1972, Dischler and Koidl 1983,
 McKenzie 1983, Duley 1984, Bussoletti et al. 1987). An extensive description and comparison of candidate amorphous carbon dust models was given by Papoular et al. \cite{pap96}.

Papoular et al. \cite{pap89} and \cite{pap01} argued that a better fit to most of the UIBs could be obtained with a model inspired by kerogens and coals. These terms cover the various forms taken by carbon-rich materials resulting from   decay, in soil, of organic matter left by living organisms near or at the earth surface (Durand et al. 1980)
. Such materials contain, therefore, a notable fraction of O, N, S, etc. besides C and H. These so-called heteroelements favor amorphousness but not completely random lattices, as may be the case with HAC.

A notable advantage of these model materials is that, over a whole century, they were chemically and physically 
analyzed by dedicated experts with up-to date instruments, and classified according to their state of natural processing, i.e. the depth at which they were extracted. Most helpful outcomes of that international effort are an abundant literature ( Durand 1980, Charcosset 1990), a vast library of spectra, the availability of samples through banks (for instance, Institut Francais du Petrole, Rueil-Malmaison, France) and a wealth of chemical models (see Spight 1994) .

To put it in a nut shell, the measured spectral properties of kerogens and coals can be reasonably reproduced by a mix of  aromatic and aliphatic small carbon structures with their peripheral hydrogen functional groups; each may have defects  such as N and S substitutions; they are mainly  linked by O atoms; varying the fractions of atoms, structures and defects helps tailoring the spectrum to fit astronomical observations. Clearly, such complex structures require a large number of atoms and so are only completed in macroscopic grains, rather than in molecules. 

As shown in the previous section and in Paper I, such grains will also carry $continuum$ and \it discrete features \rm in the visible spectral range; this is a valuable characteristic of amorphous dust models, as it may help understand IS features also observed in that range, such as the so-called Serkowski peak, a wide spectral bump which peaks in the visible and is so weak that it has only been documented in measurements of star light polarization (see Spitzer 1978, Whittet 2003).

During the last two decades, a broad range of extraterrestrial matter was isolated from meteorites and comets (e.g. Stardust mission reports in Science 2006, vol. 314), as well as from planetary ejections (e.g. from the depths of Saturn's moon, Enceladus, as reported by Postberg et al. 2018 and Khawaja et al. 2019, Cassini mission). No evidence was found of abundant large PAHs. What is seen is, rather, small, highly substituted aromatics and highly branched and O-substituted aliphatics and small organic moieties (see Cody et al. 2008), and high-mass ($>200$ a.u.) insoluble, organic, O- and N-bearing cations (Postberg et al. 2018; Bertaux and Lallement 2017).

Similarly, studying the analyzes of dust collected by the $Rosetta$ mission from comet 67P/CG, Bertaux and Lallement \cite{ber} singled out large, organic molecules found in the solid phase, which, they suggested, originated from the ISM and are responsible for DIBs. Again, no evidence for large crystal-like structures was found. It seems that the term PAH is sometimes being used now, in the astronomical literature, to designate a macromolecular material composed of aliphatic matter binding together mostly small PAHs (see Conel et al. 2008).

These recent results concur to stress the need and use of amorphous dust models.

 \end{document}